# A Straightforward Method to Judge the Completeness of a Polymorphic Gate Set

Zhifang Li, Wenjian Luo, Lihua Yue and Xufa Wang

*Abstract*—Polymorphic circuits are a special kind of circuits which possess some different build-in functions and these functions are activated by environment parameters, like light and VDD. Some theories have been proposed to guide the design of polymorphic circuits, including the definition of complete polymorphic gate sets and algorithms to judge the completeness of a polymorphic gate set.

However, the previous algorithms have to enumerate all the polymorphic signals for judging the completeness of a polymorphic gate set, and it is not easy to be conducted manually. In this paper, a straightforward method is proposed to judge the completeness of a polymorphic gate set. And the correctness of the straightforward method is proved theoretically. Some examples are given to show that the proposed method could be conducted step by step. Its actual computing cost is usually low, and it is suitable for manual operation.

## I. INTRODUCTION

Polymorphic electronic is a new field appeared in 2001 [1]. Different from the traditional electronic, the polymorphic electronic component possesses inherent build-in multiple functions. These functions are activated by the change of the environment, such as VDD, temperature, light and radiation. In different circumstances, the polymorphic electronic component performs different functions. For example, the polymorphic gate AND/OR controlled by the voltage operates as a AND gate when the power supple is 1.2V and as a OR gate when the power supple is 3.3V [1].

Recent research about polymorphic electronics focuses on fabricating polymorphic gates and the methods to build combinatorial polymorphic circuits. Some polymorphic gates have been designed and fabricated in silicon [1-5]. For example, a AND/OR gate controlled by VDD was designed in [1], and a NAND/NOR gate fabricated in a 0.7 µm CMOS was reported in [5]. Evolutionary Algorithms (EAs) have been adopted to design polymorphic circuits [4, 6-10]. In [9], a circuit which operates as a 3×4 multiplier in mode 1 and as a 7-bit sorting net in mode 2 was evolved, and it is the largest polymorphic circuit that has ever been generated. However, it is hard to generate larger scale polymorphic circuits through EAs (such as 6×6 multiplier / 12-bit sorting-net).

Zhifang Li, Wenjian Luo (Corresponding author), Lihua Yue and Xufa Wang are with the School of Computer Science and Technology, University of Science and Technology of China, Hefei 230027, China (phone:86-551-3602824). All authors are also with the Anhui Key Laboratory of Software in Computing and Communication, University of Science and Technology of China, Hefei 230027, China. Email: zhifangl@mail.ustc.edu.cn; {wjluo, llyue, xfwang}@ustc.edu.cn.

Polymorphic circuits have some potential applications, such as fault tolerance, security, adaptive systems and so on. A summary of polymorphic circuits' applications has been given in [9].

So far, there are only two papers about the completeness of the polymorphic gate set [11, 12]. In [11], the definition of a complete polymorphic gate set is given. It is proved that if the AND-Cell, OR-Cell and NOT-Cell that behave as the traditional AND, OR and NOT gates can be built by a polymorphic gate set, any polymorphic circuit can be built by this polymorphic gate set. That is to say, it is complete. In [12], deterministic algorithms are given to judge the completeness of a polymorphic gate set. It is proved that if a polymorphic gate set can build a polymorphic multiplex and the gate set in each mode is complete, the polymorphic gate set is complete.

Polymorphic gates are very different from traditional logic gates, so that the theory and methods for the traditional circuit design [13, 14] can not be applied to polymorphic circuits directly. However, in each mode, a polymorphic gate behaves just like a traditional gate. Therefore, based on the theory and designing methods for the traditional circuit, the theory and methods for polymorphic circuits can be proposed.

In this paper, based on the definition of complete polymorphic gate sets in [11], a straightforward method is proposed to judge the completeness of a polymorphic gate set. The proposed method judges whether the AND-Cell, OR-Cell and NOT-Cell can be constructed, where the AND-Cell, OR-Cell and NOT-Cell are polymorphic circuits which behave exactly the same as the traditional AND, OR and NOT logic gates in any mode [11]. This method is easy to understand, and it is suitable for manually judging the completeness of polymorphic gate sets with two or three modes. The manual version and detailed algorithm of the straightforward method are both given. The correctness of the proposed method is proved theoretically.

The rest of the paper is organized as follows. Section II introduces the related work. Section III gives the definition of polymorphic gate sets. Section IV gives the straightforward method for polymorphic gate sets with two modes. Section V proves the correctness of the method in Section IV. Section VI gives the theory and methods for polymorphic gate sets with more than two modes. Section VII gives some discussions. Finally, Section VIII concludes this paper briefly.

## II. RELATED WORK

In [11], the definition of complete polymorphic gate sets is given for the first time.

*Definition* 1. If a polymorphic gate set $P$ can construct the

AND-Cell, OR-Cell and NOT-Cell which behave as the traditional AND, OR and NOT gates in any mode, *P* is a complete polymorphic gate set [11].

Figure 1 shows an example of the AND-Cell, OR-Cell and NOT-Cell built by {AND/NOR, NAND/OR}.

In [11], it is proved that a polymorphic gate set is complete, if and only if the polymorphic gate set can construct these three cells. However, no deterministic method is given to construct the three cells.

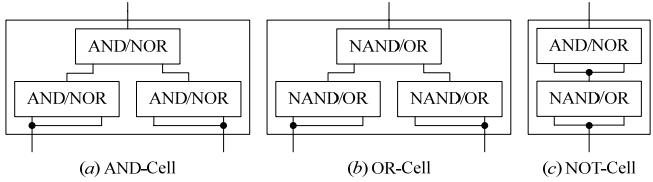

(*a*) AND-Cell    (*b*) OR-Cell    (*c*) NOT-Cell

Fig. 1. The AND-Cell, OR-Cell and NOT-Cell constructed by {AND/NOR, NAND/OR}.

In [12], it is proved that if and only if a polymorphic gate set can build a polymorphic multiplex and the gate set in each mode is complete, the polymorphic gate set is complete. Two deterministic algorithms are proposed to judge whether a polymorphic gate set can build a polymorphic multiplex. The polymorphic multiplex is a polymorphic circuit which switches different input to the output according to its working mode [6]. The design method based on the polymorphic multiplex is first proposed in [6]. Figure 2 depicts an example of a polymorphic multiplex built by {AND/NOR, NAND/OR}. In mode 1, the multiplex switches the input A to the output. In mode 2, the multiplex switches the input B to the output.

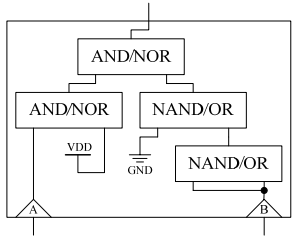

Fig. 2. The polymorphic multiplex built by {AND/NOR, NAND/OR}.

If the polymorphic multiplex can be built, and the gate set in each mode is complete, the polymorphic gate set is complete. For example, {AND/NOR, NAND/OR} can construct a polymorphic multiplex (Figure 2). Meanwhile, both of {AND, NAND} and {NOR, OR} are complete gate sets. Therefore, {AND/NOR, NAND/OR} is a complete polymorphic gate set.

In [12], the impact of logic-1 and logic-0 on the completeness of a polymorphic gate set is also discussed. Polymorphic gate sets which are complete without inputs of logic-1 and logic-0 are strong complete polymorphic gate sets. And polymorphic gate sets which are complete with inputs of logic-1 and logic-0 are weak complete polymorphic gate sets.

### III. THE DEFINITION OF THE POLYMORPHIC GATE SET

The polymorphic gate considered in this pare has two or more than two modes. For a polymorphic gate with two modes, its function in mode 1 is different from the function in mode 2. Otherwise, if it operates the same function in each mode, it is a traditional logic gate. However, as for the polymorphic gate set, the situation is complicated when the number of modes is greater than two. For example, AND/NOR/AND is a polymorphic gate with three modes. But, the function in mode 1 is the same as the function in mode 3. And obviously, {AND/NOR/AND, NOTA/AND/NOTA} is not a complete polymorphic gate set with three modes. Because both of AND/NOR/AND and NOTA/AND/NOTA performs the same function in mode 1 and mode 3, any polymorphic circuit build by the gate set would operate the same function in mode 1 and mode 3.

Therefore, in this section, the definition of a polymorphic gate set with $m$ ($m \geq 2$) modes is given. And it is shown that the Definition 1 for complete polymorphic gate sets with two modes is also correct for polymorphic gate sets with more than two modes.

***Definition 2.*** For a polymorphic gate set $P = \{p_1, \cdots, p_n\} = \{g_{1,1}/g_{1,2}/\cdots/g_{1,m}, \cdots, g_{n,1}/g_{n,2}/\cdots/g_{n,m}\}$ with $m$ modes ($m \geq 2$), $P$ satisfies that, for any $1 \leq i < j \leq m$, there exist at least one gate $p_k \in P$ and $g_{k,i} \neq g_{k,j}$.

For example, {AND/OR/NOT, XOR/OR/XOR, NAND/NOT/OR} is a polymorphic gate set with 3 modes. {AND/NAND/AND, NOT/OR/NOT} is not a polymorphic gate set with 3 modes. For any gate in {AND/NAND/AND, NOT/OR/NOT}, the function in mode 1 is the same as the function in mode 3.

In [11], for the polymorphic gate set with two modes, it is proved that if the polymorphic gate set can construct the AND-Cell, OR-Cell and NOT-Cell, the polymorphic gate set is complete. The proof is based on the multiplex, and it shows that a polymorphic gate with two modes can give the selective signal for the multiplex.

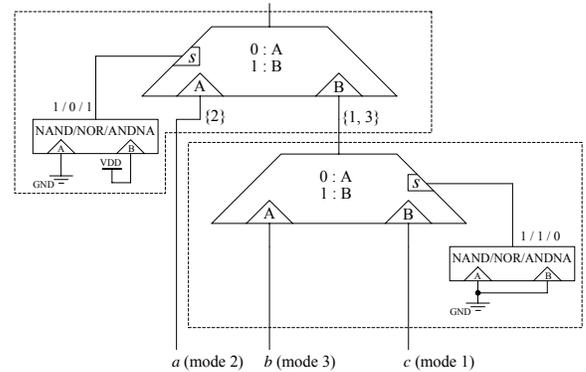

*a* (mode 2)   *b* (mode 3)   *c* (mode 1)

Fig. 3. The circuit is constructed by {NAND/NOR/ANDNA, OR/ANDNB/XOR}, and it switches different input to the output in different mode. When the selective signal *s* is 0, the multiplex switches the input at pin A to the output. When it is 1, the input at pin B is switched to the output.

In fact, the definition of complete polymorphic gate sets in [11] is also correct for polymorphic gate sets with more than two modes. For example, {NAND/NOR/ANDA, OR/ANDNB/XOR} can construct the AND-Cell, OR-Cell and NOT-Cell (the structures of NOT-Cell and AND-Cell is depicted in Figure 15 and 15, respectively). A multiplex can be built by the three cells. In Figure 3, a circuit which is constructed by the multiplex and the NAND/NOR/ANDA gate is shown. This circuit switches different input to the output in different modes. In mode 1, the input $c$ is switched to the output. In mode 2, the input $a$ is switched to the output. In mode 3, the input $b$ is switched to the output.

Because {NAND/NOR/ANDA, OR/ANDNB/XOR} can construct the AND-Cell, OR-Cell and NOT-Cell, for any function $f$, a circuit which operates the function $f$ in any mode can be constructed by the three cells. This means that for any polymorphic function $f_1/f_2/f_3$, a circuit operating as $f_1/f_2/f_3$ can be constructed in two phase [11]. 1) Three circuits which operate as $f_1$, $f_2$ and $f_3$ are built by the AND-Cell, OR-Cell and NOT-Cell, respectively. 2) The three circuits are connected to the corresponding inputs of the circuit in Figure 3. Therefore, any polymorphic function with 3 modes can be implemented by the gate set {NAND/NOR/ANDA, OR/ANDNB/XOR}.

Generally, if a polymorphic gate set $P = \{p_1, \cdots, p_n\} = \{g_{1,1}/g_{1,2}/\cdots/g_{1,m}, \cdots, g_{n,1}/g_{n,2}/\cdots/g_{n,m}\}$ ($m \geq 2$) can construct the AND-Cell, OR-Cell and NOT-Cell, the polymorphic gate set is complete. Firstly, it is shown that a polymorphic circuit with $m$ inputs and one output can be built, and this circuit switches different input to the output in different modes. Figure 4 shows the structure of such a circuit. And the circuit can be built in following steps. In fact, this method is similar the approach in [11], i.e. in each step, it is shown that the selective signal can be constructed.

Firstly, a multiplex is built by the AND-Cell, OR-Cell and NOT-Cell. The symbol $s$ is the selective signal, and the two input pins are labeled as A and B. When $s$ is 0, the input at pin A is switched to the output. When $s$ is 1, the input at pin B is switched to the output.

Secondly, according to Definition 2, there exist a polymorphic gate $p_{k_1} \in P$ and $g_{k_1,1} \neq g_{k_1,2}$. Therefore, there exist at least one combination of inputs $(in_1, in_2) \in \{00, 01, 10, 11\}$ and $p_k$ outputs differently in mode 1 and 2 with $(in_1, in_2)$ as its input. Without the loss of generality, suppose $p_k$ outputs 0 in modes $\{i_1, i_2, \cdots, i_u\}$ ($1 = i_1 < i_2 < \cdots < i_u \leq m$) and outputs 1 in modes $\{j_1, j_2, \cdots, j_v\}$ ($2 = j_1 < j_2 < \cdots < j_v \leq m$), where $\{i_1, i_2, \cdots, i_u\} \cap \{j_1, j_2, \cdots, j_v\} = \phi$, $0 < u, v < m$ and $u + v = m$.

When $p_{k_1}$ is connected to the selective signal $s$ of the multiplex, the input at pin A is switched to the output in modes $\{i_1, i_2, \cdots, i_u\}$, and the input at pin B is switched to the output in modes $\{j_1, j_2, \cdots, j_v\}$. The dashed rectangle labeled "I" in Figure 4 shows the combination of $p_{k_1}$ and the multiplex. In this step, the modes are separated to two sets $\{i_1, i_2, \cdots, i_u\}$ and $\{j_1, j_2, \cdots, j_v\}$.

Thirdly, as for $\{i_1, i_2, \cdots, i_u\}$, if its size is one, the process stops. Otherwise, according to Definition 2, there exist a polymorphic gate $p_{k_2} \in P$ and $g_{k_2,i_1} \neq g_{k_2,i_2}$. Similar to the analysis in step 2, the combination of $p_{k_2}$ and the multiplex forms a circuit depicted in the dashed rectangle labeled "II" in Figure 4, and this circuit separate $\{i_1, i_2, \cdots, i_u\}$ to $\{i_{A,1}, i_{A,2}, \cdots, i_{A,r}\}$ and $\{i_{B,1}, i_{B,2}, \cdots, i_{B,t}\}$, where $\{i_{A,1}, i_{A,2}, \cdots, i_{A,r}\} \cap \{i_{B,1}, i_{B,2}, \cdots, i_{B,t}\} = \phi$, $0 < r, t < u$ and $r + t = u$. In mode $\{i_{A,1}, i_{A,2}, \cdots, i_{A,r}\}$, the input at pin A is switched to the output. In mode $\{i_{B,1}, i_{B,2}, \cdots, i_{B,t}\}$, the input at pin B is switched to the output. The same analysis can be adopted to $\{j_1, j_2, \cdots, j_v\}$, and the circuit in the dashed rectangle labeled as "III" in Figure 4 is obtained.

This process is adopted, and finally the polymorphic circuit wanted is obtained. Figure 3 is an example of such a polymorphic circuit.

For any polymorphic function $f_1/f_2/\cdots/f_m$ with $m$ modes, $m$ circuits, which perform functions $f_1, f_2, \cdots, f_m$, respectively, can be built by AND-Cell, OR-Cell and NOT-Cell. These $m$ circuits are connected to the corresponding inputs of the circuit in Figure 4, and a polymorphic circuit which performs the function $f_1/f_2/\cdots/f_m$ is obtained. So, if a polymorphic gate set can build the AND-Cell, OR-Cell and NOT-Cell, it is complete.

IV. THE STRAIGHTFORWARD METHOD TO JUDGE THE COMPLETENESS OF POLYMORPHIC GATE SETS WITH TWO MODES

The straightforward method judges the completeness of a polymorphic gate set through building the AND-Cell, OR-Cell and NOT-Cell. Here, the AND-Cell is taken as an example, in the first step, some polymorphic circuits which perform the AND function in mode 1 are constructed. In the second step, those polymorphic circuits obtained in the first step are adopted to construct a circuit which perform the AND function in mode 2. If the above two steps can be accomplished successfully, the AND-Cell is obtained. The straightforward method is described in Section 3.1.

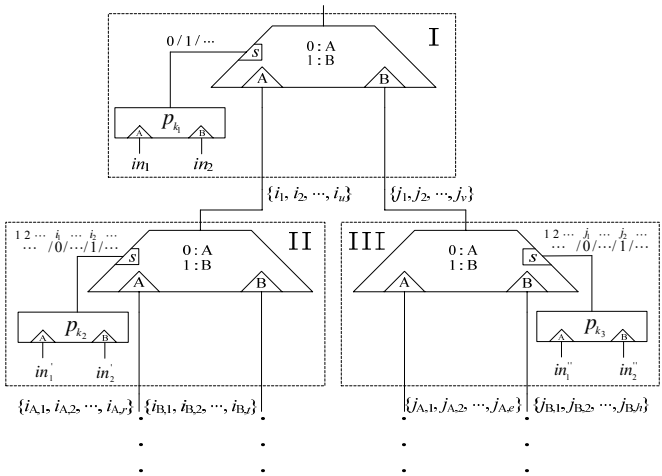

Fig. 4. The circuit which switches different input to the output in different mode.

For convenience, firstly, the symbols and functions of the 16 types of traditional logic gates are shown in Table I. This table is from [12]. In this paper, under each working mode, a polymorphic gate would operate as one of traditional logic gates in Table I. In the rest of this paper, $p(a, b)$ denotes the gate $p$ with $a$ and $b$ as its inputs, where $a$ is fed to the input pin A of $p$ and $b$ is fed to the input pin $B$. In fact, NOTA, NOTB, WIREA and WIREB are logic gates with one input. Therefore, sometimes, the symbol WIRE is adopted to stand for the function WIREA and WIREB, and the symbol NOT is adopted to stand for the function NOTA and NOTB.

TABLE I. The 16 types of traditional combinational logic gates. Symbols "A" and "B" denote the two inputs of each gate, respectively.

| Symbol | AND | OR | NAND | NOR | ANDNA | ANDNB |
|---|---|---|---|---|---|---|
| Function | $AB$ | $A+B$ | $\overline{AB}$ | $\overline{A+B}$ | $\overline{A}B$ | $A\overline{B}$ |
| Symbol | XOR | NXOR | NOTA | NOTB | WIREA | WIREB |
| Function | $A \oplus B$ | $\overline{A \oplus B}$ | $\overline{A}$ | $\overline{B}$ | $A$ | $B$ |
| Symbol | ORNA | ORNB | ZERO | ONE | | |
| Function | $\overline{A}+B$ | $A+\overline{B}$ | $0$ | $1$ | | |

### A. The Straightforward Method

In this subsection, the detailed process to construct the AND-Cell manually is presented, and the process to construct the OR-Cell and NOT-Cell is similar to the AND-Cell.

Suppose a polymorphic gate set $P = \{p_1, \cdots, p_n\} = \{g_{1,1}/g_{1,2}, \cdots, g_{n,1}/g_{n,2}\}$ is given, and $p_i = g_{i,1}/g_{i,2}$ $(1 \leq i \leq n)$.

**Step 1.** Firstly, $\{g_{1,1}, g_{2,1}, \cdots, g_{n,1}, \text{logic-1}, \text{logic-0}\}$ is adopted to construct circuits $C_1, C_2, \cdots, C_k$. Each $C_i$ $(1 \leq i \leq k)$ performs the AND function, and for any $1 \leq i < j \leq k$, $C_i$ and $C_j$ are different from each other. Then, for each $g_{i,1}$ $(1 \leq i \leq n)$ that is a building block of $C_j$, $g_{i,1}$ is replaced by $p_i$. So, a polymorphic circuit $PC_j$ is obtained which operates as a AND gate in mode 1. Therefore, a set of polymorphic circuits $P' = \{PC_1, \cdots, PC_k\} = \{p'_1, \cdots, p'_k\} = \{\text{AND}/g'_{1,2}, \cdots, \text{AND}/g'_{k,2}\}$ is obtained. Obviously, each $g'_{i,2}$ operates as one of the 16 traditional logic gates.

**Step 2.** Check whether $\{g'_{1,2}, g'_{2,2}, \cdots, g'_{k,2}, \text{logic-1}\}$ can construct a circuit $Cir$ which operates as a AND gate.

1) If such a circuit $Cir$ operating as a AND gate can be constructed by $\{g'_{1,2}, g'_{2,2}, \cdots, g'_{k,2}, \text{logic-1}\}$, each gate $g'_{i,2}$ $(1 \leq i \leq k)$ that is a building block of $Cir$ is replaced by $p'_i$, and the new obtained polymorphic circuit is denoted as $Cir_{pol}$. In mode 2, $Cir_{pol}$ would operate as a AND gate. In Section 4, it is proved that $Cir_{pol}$ would also operate as AND in mode 1. In other words, $P$ can construct the AND-Cell.

2) If $\{g'_{1,2}, g'_{2,2}, \cdots, g'_{k,2}, \text{logic-1}\}$ can not construct a AND gate, it does not mean that $P$ can not construct the AND-Cell. It is possible that $P$ could construct another polymorphic gate AND/$g$ that is different from any gate in $P'$, and $\{g'_{1,2}, g'_{2,2}, \cdots, g'_{k,2}, g, \text{logic-1}\}$ could construct a circuit which operates as a AND gate. Therefore, we should go back to step 1 to check whether some other circuits, which are different from $C_1, C_2, \cdots, C_k$ and operate as a AND gate, could be generated by $\{g_{1,1}, g_{2,1}, \cdots, g_{n,1}, \text{logic-1}, \text{logic-0}\}$.

If such circuits can be generated by $\{g_{1,1}, g_{2,1}, \cdots, g_{n,1}, \text{logic-1}, \text{logic-0}\}$ and denoted as $C_{k+1}, C_{k+2}, \cdots, C_{k+t}$, polymorphic circuits $\{PC_{k+1}, \cdots, PC_{k+t}\} = \{\text{AND}/g'_{k+1,2}, \cdots, \text{AND}/g'_{k+t,2}\}$ can be obtained from $\{C_{k+1}, C_{k+2}, \cdots, C_{k+t}\}$ by replacing every gate $g_{i,1}$ $(1 \leq i \leq n)$ of $C_j$ $(k < j \leq k+t)$ with $p_i$. If $\{g'_{k+1,2}, \cdots, g'_{k+t,2}\} \not\subset \{g'_{1,2}, \cdots, g'_{k,2}\}$, we go to step 2 with $\{g'_{k+1,2}, \cdots, g'_{k+t,2}\} \cup \{g'_{1,2}, \cdots, g'_{k,2}, \text{logic-1}\}$. Otherwise, if $\{g'_{k+1,2}, \cdots, g'_{k+t,2}\} \subseteq \{g'_{1,2}, \cdots, g'_{k,2}\}$, we go back to step 1 to see whether some circuits different from $\{C_1, C_2, \cdots, C_k, \cdots, C_{k+t}\}$ can be built, or we terminate the process.

In the above method, the AND-Cell is taken as an example. Although the two above steps are easy to be carried out manually, it is hard to enumerate all the combinations of those gates. For a polymorphic gate set, if the process gets stuck at step 1 or step 2, it does not make sure that the polymorphic gate set can not construct the AND-Cell. Therefore, the algorithms are given in Figure 5 and Figure 6 to judge whether the AND-Cell, OR-Cell and NOT-Cell can be constructed by a polymorphic gate set. The algorithm can explore all possible structures of circuits operating as a polymorphic gate, and gives a deterministic result whether the AND-Cell (or OR-Cell or NOT-Cell) can be generated.

---

$Judge(P, f)$
in: A polymorphic gate set $P = \{p_1, p_2, \cdots, p_n\} = \{g_{1,1}/g_{1,2}, g_{2,1}/g_{2,2}, \cdots, g_{n,1}/g_{n,2}\}$ and $p_i = g_{i,1}/g_{i,2}$ $(1 \leq i \leq n)$. The parameter $f$ is AND or OR or NOT.
out: If $P$ can construct a circuit which operates the function $f$ in any mode, return true. Otherwise, return false.

1. $d \leftarrow 1$
2. $R \leftarrow \phi, I \leftarrow \phi$
3. *stack* is initialized to empty
4. **do**{
5.   **if** $f \neq$ NOT **then**
6.     $\{P\_new, R, I\} \leftarrow Construction\_ANDOR(P, f, d, R, I, 2)$
7.   **else**
8.     $\{P\_new, R, I\} \leftarrow Construction\_NOT(P, d, R, I, 2)$
9.   **if** $R = \phi$ and $d = 1$ **then**
10.     **return** false
11.   **else**
12.     **if** $R = \phi$ **then**
13.       $d \leftarrow d - 1$
14.       $\{P, R, I\} \leftarrow stack.pop()$
15.     **else**
16.       $d \leftarrow d + 1$
17.       $stack.push(\{P\_new, R, I\})$
18.       $P \leftarrow R, R \leftarrow \phi, I \leftarrow \phi$
19. }**while**($d < 2$ or $R = \phi$)
20. **return** true

Fig. 5. $Judge(P, f)$ judges whether a polymorphic gate set $P$ can implement the function $f$. All polymorphic gates generated by $P$ at line 5 form the set $I$. At line 5, those polymorphic gates operating the function $f$ in the first $d$ modes form the set $R$.

In Figure 5, if *Judge*(*P*, *f*) returns true, *P* can construct the circuit which operates the function *f* in any mode. Otherwise, if it returns false, *P* can not build such polymorphic circuits. When the parameter *d* is 1 and 2, the operation at line 6 or line 8 corresponds to the **Step 1** and **Step 2** in the manual operation, respectively.

When the parameter *d* is 1, (i) if the set *R* is not empty after the operation at line 6 or line 8, some polymorphic circuits which operate the function *f* in mode 1 have been built. Therefore, *d* is increased by 1, parameters {*P*, *R*, *I*} are stored in *stack*, and *R* is set to the polymorphic gate set *P* for the next loop (line 16 to line 18). (ii) If *R* is empty (line 9), the polymorphic gate set can not build the function *f*, so that the algorithm would terminate and return false.

When the parameter *d* is 2, (i) if *R* is not empty after the operation at line 5, it means a polymorphic circuit which operates the function *f* both in mode 1 and mode 2 has been constructed, and the algorithm would terminate and return true. (ii) If *R* is empty (line 12), it means that the current polymorphic gate set *P* obtained in the last loop can not build the circuit which operate the function *f* both in mode 1 and mode 2. Therefore, the {*P*, *R*, *I*} are popped from the stack (line 14), and in the next loop, some new polymorphic circuits which operate the function *f* in mode 1 would be built.

---

*Construction_ANDOR*(*P*, *f*, *d*, *R*, *I*, *m*)

in:  A polymorphic gate set $P = \{p_1, p_2, \cdots, p_n\} = \{g_{1,1}/g_{1,2}/\cdots/g_{1,m}, \cdots, g_{n,1}/g_{n,2}/\cdots/g_{n,m}\}$. The parameter *f* is AND or OR. For any $1 \le i < d$ ($d \ge 2$) and $1 \le j \le n$, $g_{j,i} = f$. *d* is an integer and $1 \le d \le m$. *P_new* is the union of the polymorphic gate set *P* and the new generated gate set *I_new*.
If $d = 1$ then $S \leftarrow \{0, 1, a, b\}$.
If $d > 1$ and $f =$ AND then $S \leftarrow \{1, a, b\}$
If $d > 1$ and $f =$ OR then $S \leftarrow \{0, a, b\}$

out: {*P_new*, *R*, *I*}

1. *P_new* ← *P* ∪ {VA, VB}
2. *h* ← |*R*|
3. *loop* ← 0
4. **do** {
5.   *loop* ← *loop* + 1
6.   *I_new* ← ϕ
7.   **for** every ($pol_1$, $pol_2$, $pol_3$) ∈ *P_new*×*P_new*×*P_new*, $pol_1 \notin$ {VA, VB} and $u, v, w, t \in S$ **do**
8.     *pol* ← $pol_1$( $pol_2(u, v)$, $pol_3(w, t)$ )
9.     **if** *pol* ∉ *I* ∪ *I_new* **then** *I_new* ← *I_new* ∪ {*pol*}
10.    **if** *pol*[*d*] = *f* and *pol* ∉ *R* **then** *R* ← *R* ∪ {*pol*}
11.   *I* ← *I* ∪ *I_new*
12.   *P_new* ← *P_new* ∪ *I_new*
13.   **if** *I_new* ≠ ϕ and |*R*| > *h* **then return** {*P_new*, *R*, *I*}
14. } **while**(*I_new* ≠ ϕ)
15. **return** {ϕ, ϕ}

Fig. 6. The subroutine constructs the AND (OR) function under mode *d* with polymorphic gate set *P*. VA and VB are two virtual polymorphic gates which operate the function WIREA and WIREB in any mode, respectively. In fact, VA (VB) is a wire connected to the input pin A (B). *pol* is a polymorphic gate, and *pol*[*d*] denotes the function that *pol* operates in mode *d*. *loop* is an integer variable which records the iterations of the Do-While loop, and it is used for the analysis of the algorithm's time complexity.

The procedure *Construction_ANDOR*(*P*, *f*, *d*, *R*, *I*, *m*) in Figure 6 builds some polymorphic circuits, and all these circuits operate the function *f* (*f* ∈ {AND, OR}) in mode *d*. In Figure 6, if the parameter *d* is 1, the logic-0 and logic-1 can be used in the construction of the polymorphic gate set *R*. If *d* is greater than 1 and *f* is AND, logic-1 can be used. If *d* is greater than 1 and *f* is OR, logic-0 can be used.

The procedure *Construction_NOT*(*P*, *d*, *R*, *I*, *m*) in Figure 7 builds some polymorphic circuits which operate the function NOT in mode *d*.

---

*Construction_NOT*(*P*, *d*, *R*, *I*, *m*)

in:  A polymorphic gate set $P = \{p_1, p_2, \cdots, p_n\} = \{g_{1,1}/g_{1,2}/\cdots/g_{1,m}, \cdots, g_{n,1}/g_{n,2}/\cdots/g_{n,m}\}$. For any $1 \le i < d$ ($d \ge 2$) and $1 \le j \le n$, $g_{j,i} = f$. *d* is an integer and $1 \le d \le m$. *P_new* is the union of the polymorphic gate set *P* and the new generated gate set *I_new*.

out: {*P_new*, *R*, *I*}

1. *P_new* ← *P* ∪ {VA, VB}
2. *h* ← |*R*|
3. *loop* ← 0
4. **do** {
5.   *loop* ← *loop* + 1
6.   *I_new* ← ϕ
7.   **if** *d* = 1 **then**
8.     **for** every ($pol_1$, $pol_2$, $pol_3$) ∈ *P_new*×*P_new*×*P_new*, $pol_1 \notin$ {VA, VB} and $u, v, w, t \in \{0, 1, a\}$ **do**
9.       *pol* ← $pol_1$( $pol_2(u, v)$, $pol_3(w, t)$ )
10.      **if** *pol* ∉ *I* ∪ *I_new* **then** *I_new* ← *I_new* ∪ {*pol*}
11.      **if** *pol*[1] = *f* and *pol* ∉ *R* **then** *R* ← *R* ∪ {*pol*}
12.   **else**
13.     **for** every ($pol_1$, $pol_2$) ∈ *P_new*×*P_new* and $pol_1 \notin$ {VA,VB} **do**
14.       *pol* ← $pol_1$( $pol_2(a)$)
15.      **if** *pol* ∉ *I* ∪ *I_new* **then** *I_new* ← *I_new* ∪ {*pol*}
16.      **if** *pol*[*d*] = *f* and *pol*[1] = *f* and *pol* ∉ *R* **then** *R* ← *R* ∪ {*pol*}
17.   *I* ← *I* ∪ *I_new*
18.   *P_new* ← *P* ∪ *I_new*
19.   **if** *I_new* ≠ ϕ and |*R*| > *h* **then return** {*P_new*, *R*, *I*}
20. } **while**(*I_new* ≠ ϕ)
21. **return** {ϕ, ϕ}

Fig. 7. The subroutine constructs the NOT function under mode *d* with the polymorphic gate set *P*. The symbol VA, VB, *pol*[*d*] and *loop* are the same as those in Figure 6.

---

In fact, for any polymorphic gate which operates the function *f* in the first *d* modes, if *P* can construct this polymorphic gate, the "*Construction_ANDOR*(*P*, *f*, *d*, …)" can generate this gate.

Suppose a polymorphic circuit *C* is built by *P*, where *C* operates as a polymorphic gate and it performs the function *f* in the first *d* modes. The gate that gives the output of *C* is *p* (*p* ∈ *P*), and two subcircuits connected to *p* are $C_A$ and $C_B$, respectively. Assume "*Construction_ANDOR*(*P*, *f*, *d*, …)" can not construct a polymorphic gate which operates as *C*. That is to say, "*Construction_ANDOR*(*P*, *f*, *d*, …)" can not build a polymorphic gate operating as $C_A$ or $C_B$. Otherwise, if two polymorphic gates $p_A$ and $p_B$, which operate as $C_A$ and $C_B$, respectively, can be built by "*Construction_ANDOR*(*P*, *f*, *d*, …)", in the next Do-While loop (stated at line 4), when $pol_1$, $pol_2$ and $pol_3$ are *p*, $p_A$ and $p_B$, respectively, *p*( $p_A(\ldots)$, $p_B(\ldots)$ ) can be obtained (line 8) and it operates as *C*. This contradicts with the assumption that *C* can not be built by "*Construction_ANDOR*(*P*, *f*, *d*, …)". Therefore, at least one of $C_A$ and $C_B$ can not be built by "*Construction_ANDOR*(*P*, *f*, *d*, …)". Suppose $C_A$ can not be built, and the gate that gives the output of $C_A$ is $p'$. The two subcircuits connected to $p'$ are $C_{AA}$ and $C_{AB}$. Similar to the analysis of $C_A$ and $C_B$, at least one of $C_{AA}$ and $C_{AB}$ can not be built by "*Construction_ANDOR(P, f, d, …)*".

This induction is applied until a gate is met, and the inputs of the gate are the inputs of $C$. Obviously, this gate belongs to $P$. However, according to the analysis, this gate can not be built by "*Construction_ANDOR(P, f, d, …)*". This contradicts that this gate belongs to $P$. Therefore, the assumption is invalid. That is to say, if a polymorphic gate which operates the function $f$ ($f \in$ {AND, OR}) in the first $d$ modes can be constructed by $P$, it can be constructed by "*Construction_ANDOR(P, f, d, …)*".

Similarly, for any polymorphic gate which operates the function NOT in the first $d$ modes, if the polymorphic gate set $P$ can build it, "*Construction_NOT(P, d, …)*" can generate it.

On the other hand, if a polymorphic gate can not be constructed by $P$, "*Construction_NOT(P, d, …)*" and "*Construction_ANDOR(P, f, d, …)*" also can not build it.

So, if $P$ can construct a polymorphic circuit which operates the function $f$ both in mode 1 and mode 2, "*judge(P, f)*" returns true. Otherwise, it returns false.

*B. Some Examples*

In this subsection, complete polymorphic gate sets {NAND/NOR}, {AND/NOTA, NOTA/OR} and {NOR/XOR, XOR/NAND} are adopted as examples to show the process of the manual method.

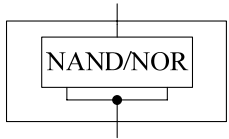

Fig. 8. The NOT-Cell built by NAND/NOR

**Example 3-1 {NAND/NOR}**

Obviously, the NOT-Cell can be obtained from {NAND/NOR} easily, which is shown in Figure 8. The procedure of constructing the AND-Cell is given as follows.

Step 1: Firstly, some polymorphic gates which operate the AND function in mode 1 are built.

1) Because NOT(NAND($a$, $b$)) = AND($a$, $b$), the combination of NOT-Cell and NAND/NOR could generate AND/OR, which is depicted in Figure 9(*a*).

2) If a polymorphic gate which is different from NOT-Cell and operates the NOT function in mode 1 can be built, the combination of this gate and NAND/NOR could operate AND function in mode 1 and operate another function different from OR in mode 2. Fortunately, a NOTB/ANDNB gate can be built from NAND/NOR and ONE/NOT, as is shown in Figure 9(*b*). The ONE/NOT gate is given in Figure 9(*c*). Therefore, a polymorphic gate AND/ANDNA can be built from NOTB/ANDNB, NAND/NOR and NOT-Cell, as is shown in Figure 9(*d*).

Finally, the polymorphic circuit set {AND/OR, AND/ANDNA} is obtained.

Step 2: {OR, ANDNA} can build a circuit operating the AND function as depicted in Figure 9(*e*). Therefore, the AND-Cell can be built by {NAND/NOR}. The structure of the AND-Cell is depicted in Figure 9(*f*).

According to De-Morgan rules, the OR-Cell can be obtained from NOT-Cell and AND-Cell. So, according to Definition 1, {NAND/NOR} is a complete polymorphic gate set.

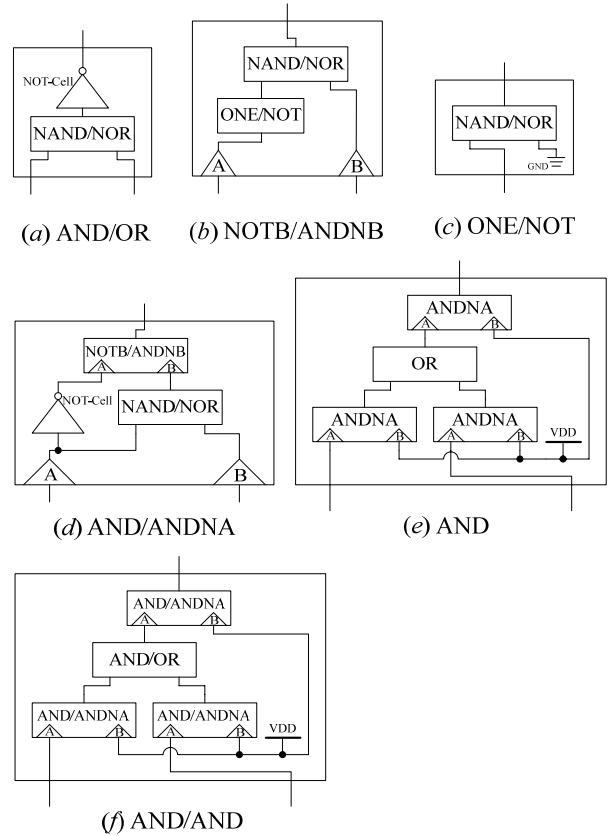

(*a*) AND/OR  (*b*) NOTB/ANDNB  (*c*) ONE/NOT

(*d*) AND/ANDNA  (*e*) AND

(*f*) AND/AND

Fig. 9. (*a* ~ *d*) Polymorphic gates built in the Step 1. (*e*) The structure of the AND gate built by ANDNA and OR. (*f*) The structure of the AND-Cell

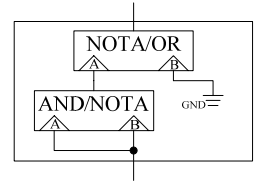

Fig. 10. The NOT-Cell built by {NOTA/OR, AND/NOTA}.

**Example 3-2 {AND/NOTA, NOTA/OR}**

It is easy to build the NOT-Cell and its structure is depicted in Figure 10. As for the AND-Cell, in the first step, AND/ORNA can be obtained and its structure is depicted in Figure 11(*a*). Because {NOT, ORNA} can build a AND gate shown in Figure 11(*b*), {AND/NOTA, AND/ORNA} can build the AND-Cell depicted in Figure 11(*c*).

According to De-Morgan rules, the OR-Cell can be obtained from NOT-Cell and AND-Cell. Therefore, according to Definition 1, {AND/NOTA, AND/ORNA} is a complete polymorphic gate set.

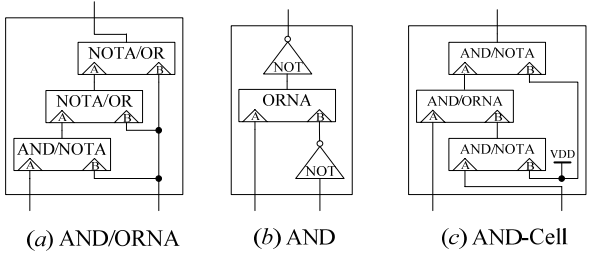

Fig. 11. (*a*) the structure of AND/ORNA built by {AND/NOTA, NOTA/OR}. (*b*) {NOT, ORNA} can built a AND gate. (*c*) The structure of the AND-Cell.

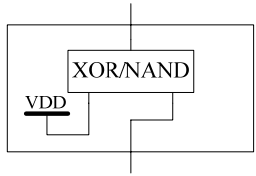

Fig. 12. The NOT-Cell built by XOR/NAND

**Example 3-3 {NOR/XOR, XOR/NAND}**

Because XOR(1, $a$) = NOTB($a$) and NAND(1, $a$) = NOTB($a$), the NOT-Cell can be built by XOR/NAND as that depicted in Figure 12.

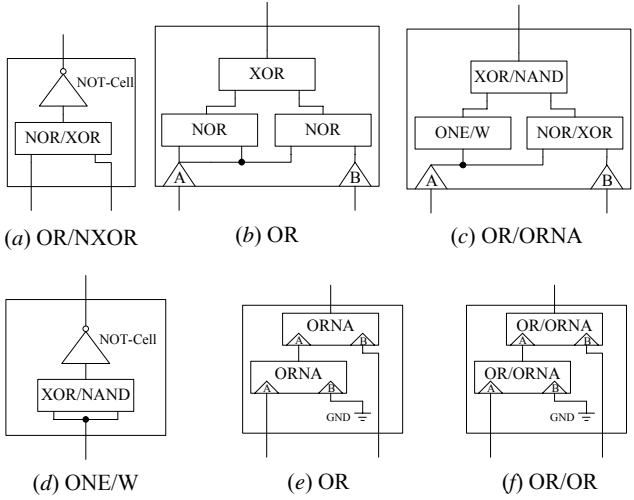

Fig. 13. The construction of the OR-Cell.

The process to build the OR-Cell is given as follows.

Step 1. As shown in Figure 13(*a*), the combination of NOT-Cell and NOR/XOR can generate OR/NXOR.

{XOR, NOR} can build an OR gate as depicted in Figure 13(*b*). In Figure 13(*c*), an OR/ORNA polymorphic gate can be obtained from Figure 13(*a*) by replacing the XOR with XOR/NAND, NOR with NOR/XOR and the ONE with ONE/W. The ONE/W polymorphic gate can be built as in Figure 13(*d*).

Finally, the polymorphic gate set {OR/NXOR, OR/ORNA} is obtained.

Step 2. {NXOR, ORNA} can build an OR gate shown in Figure 13(*e*). Therefore, {NOR/XOR, XOR/NAND} can build the OR-Cell, which is depicted in Figure 13(*f*).

According to De-Morgan rules, the AND-Cell can be built by the OR-Cell and NOT-Cell. So, {NOR/XOR, XOR/NAND} a complete polymorphic gate set.

It can be observed from these examples that the NOT-Cell plays an important part in constructing the other two cells. Usually, it is easier to build the NOT-Cell. Therefore, it is recommended that the NOT-Cell should be constructed before the AND-Cell and OR-Cell.

V. PROOFS OF THE STRAIGHTFORWARD METHOD

Firstly, for convenience, the definition of the depth of a circuit is given.

***Definition* 3.** Suppose $C$ is a one-output circuit. When a signal is fed into inputs of $C$, the largest number of gates that the signal passes from the input to the output is the depth of the circuit $C$ [12].

**LEMMA 1.** Suppose $C$ is a circuit with one output and two inputs. For each gate $g$ which is a building block of $C$, neither input pin A nor input pin B of $g$ is connected to logic-0. If each gate of $C$ is replaced with a AND gate, the new obtained circuit $C'$ would operate as a AND gate.

**PROOF.** This lemma is proved by the induction on the depth $H$ of the circuit $C$. The two inputs of $C$ are denoted as $a$ and $b$, respectively. $C'$ is a circuit obtained by replacing each gate of $C$ with a AND gate.

(1) $H = 1$. $C$ is a circuit composed by one gate. Because AND($a$, $b$) = $ab$, $C'$ would operate as a AND gate. Therefore, the statement is true when $H = 1$.

(2) For the induction step, assume the statement is true when $H \leq n$.

(3) $H = n + 1$. Suppose $g_o$ is the gate which gives the output of $C$, $C_A$ is the subcircuit which is connected to the input pin A of $g_o$, and $C_B$ is the subcircuit which is connected to the input pin B of $g_o$. Because the depth of $C$ is $n + 1$, depths of $C_A$ and $C_B$ are less than $n + 1$. $C'_A$ and $C'_B$ are two circuits obtained by replacing every gate of $C_A$ and $C_B$ with AND gates, respectively.

3.1) The inputs of $C_A$ is {$a$, $b$, logic-1}. According to the induction assumption, $C'_A$ would operate as AND($a$, $b$) = $ab$.

3.1.1) The inputs of $C_B$ is {$a$, $b$, logic-1}. $C'_B$ would operate as AND($a$, $b$) = $ab$. Because AND($ab$, $ab$) = $ab$, $C'$ would operate as a AND gate.

3.1.2) The inputs of $C_B$ is {$a$, logic-1}. Therefore, the number of inputs of $C'_B$ is one and it is $a$. According to induction assumption, for any circuit $Cir$, if its depth is less than $n + 1$, any gate of $Cir$ is a AND gate and the inputs of $Cir$ is {*a*, *b*, logic-1}, *Cir* would operate as AND(*a*, *b*). When *b* is logic-1 or *a*, *Cir* would operate as AND(*a*, 1) = WIRE(*a*), or AND(*a*, *a*) = WIRE(*a*). Therefore, $C_B^{'}$ would operate as a WIRE gate.

3.1.3) The inputs of $C_B$ is {logic-1}. That is to say, for each gate *g* of $C_B$, the inputs of *g* is either connected to logic-1 or another gate of $C_B$. Similar to the analysis in 3.1.2), $C_B^{'}$ would operate as a ONE gate.

3.2) The number of inputs of $C_A$ is {*a*, logic-1}. According to the analysis in 3.1.2), $C_A^{'}$ would operate as a WIRE gate. Let's denote the input of $C_A$ as *a*.

3.2.1) The number of inputs of $C_B$ is {*a*, *b*, logic-1}. Similar to 3.1.2).

3.2.2) The number of inputs of $C_B$ is {*b*, logic-1}. According to the analysis of 3.1.2), $C_B^{'}$ would operate as WIRE(*b*) = *b*. Therefore, $C^{'}$ would operate as AND(WIRE(*a*), WIRE(*b*)) = *ab*.

In summary, the statement is true when $H = n + 1$. ∎

A polymorphic gate set $P = \{p_1, p_2, ..., p_n\} = \{AND/g_{1,2}, AND/g_{2,2}, ..., AND/g_{n,2}\}$, and $p_i = AND/g_{i,2}$ ($1 \le i \le n$). Suppose *C* is a circuit built by $\{g_{1,2}, g_{2,2}, ..., g_{n,2}, \text{logic-1}\}$, and it performs the function AND. If each gate $g_{j,2} \in C$ ($1 \le j \le n$) is replaced by $p_j$, the new obtained polymorphic circuit is denoted as $C_{pol}$. Lemma 1 indicates that $C_{pol}$ would operate the AND function in mode 1, i.e. $C_{pol}$ is a AND-Cell.

**LEMMA 2.** Suppose *C* is a circuit with one output and two inputs, and for each gate *g* that is a building block of *C*, neither input pin A nor input pin B of *g* is connected to logic-1. If each gate of *C* is replaced with an OR gate, the new obtained circuit $C^{'}$ would operate as an OR gate.

Lemma 2 can be proved by the way similar to Lemma 1. In the following, let's denote *tgate* as the gate set {AND, OR, NAND, NOR, ANDNA, ORNA, ORNB, XOR, NXOR, NOTA, NOTB, WIREA, WIREB, ZERO, ONE}.

**LEMMA 3.** A polymorphic gate set $P = \{p_1, p_2, ..., p_n\} = \{g_{1,1}/g_{1,2}, g_{2,1}/g_{2,2}, ..., g_{n,1}/g_{n,2}\}$, and $p_i = g_{i,1}/g_{i,2}$ ($1 \le i \le n$). Each $p_i$ performs the function $g_{i,1}$ in mode 1 and performs the function $g_{i,2}$ in mode 2. If *P* is complete, *P* can construct another polymorphic gate set $P^{'} = \{p_1^{'}, p_2^{'}, ..., p_k^{'}\} = \{AND/g_1, AND/g_2, ..., AND/g_k\}$, and $P^{'}$ can construct a special polymorphic circuit which operates as AND/AND.

**PROOF.** Obviously, there exist a gate set $G = \{g_1, g_2, ..., g_k\}$, $G \subseteq tgate$ and $G \cup \{\text{logic-1}\}$ can construct a circuit *C* operating as a AND gate. For example, because NOTA(OR(NOTA(*a*), NOTA(*b*))) = AND(*a*, *b*), {NOTA, OR} can construct a AND gate. Because ANDNA(ANDNA(*a*, 1), *b*) = AND(*a*, *b*), {ANDNA, logic-1} can construct a AND gate.

Because *P* is a complete polymorphic gate set, the polymorphic circuit AND/$g_i$ can be built by *P* for each $g_i \in G$ ($1 \le i \le 16$). In other words, *P* can construct a polymorphic gate set $P^{'} = \{AND/g_1, AND/g_2, ..., AND/g_k\}$. Therefore, a polymorphic circuit $C_{pol}^{'}$ can be obtained by replacing every $g_i \in C$ with AND/$g_i$ for each $1 \le i \le k$. Because the logic-0 is never adopted in the construction of *C*, according to Lemma 1, in mode 1, $C_{pol}^{'}$ would operate as a AND gate. So, $C_{pol}^{'}$ would operate as a AND/AND gate. ∎

In fact, Lemma 3 indicates that an AND-Cell could be built by a polymorphic gate set in two independent steps, i.e. the straightforward method proposed in Section 3.

In the following, Lemma 4 and Lemma 5 indicate the similar consequence about OR-Cell and NOT-Cell. Therefore, Lemma 3, Lemma 4 and Lemma 5 prove the correctness of the straightforward method proposed in Section 3.

**LEMMA 4.** A polymorphic gate set $P = \{p_1, p_2, ..., p_n\} = \{g_{1,1}/g_{1,2}, g_{2,1}/g_{2,2}, ..., g_{n,1}/g_{n,2}\}$ and $p_i = g_{i,1}/g_{i,2}$ ($1 \le i \le n$). If *P* is complete, *P* can construct another polymorphic gate set $P^{'} = \{p_1^{'}, p_2^{'}, ..., p_k^{'}\} = \{OR/g_1, OR/g_2, ..., OR/g_k\}$, and $P^{'}$ can construct a polymorphic circuit which operates as OR/OR.

Lemma 4 can be proved by the way similar to Lemma 3. Lemma 3 and Lemma 4 indicate that if a polymorphic gate set is complete, the straightforward method can construct the AND-Cell and OR-Cell.

**LEMMA 5.** A polymorphic gate set $P = \{p_1, p_2, ..., p_n\} = \{g_{1,1}/g_{1,2}, g_{2,1}/g_{2,2}, ..., g_{n,1}/g_{n,2}\}$ and $p_i = g_{i,1}/g_{i,2}$ ($1 \le i \le n$). If *P* is complete, the straightforward method can build NOT/NOT.

**PROOF.** In the construction of NOT/NOT, only {0, 1, *a*} is adopted as the input, so that the polymorphic gate set $P^{'}$ obtained in the first step satisfies that $P^{'} \subseteq \{\text{NOT/ZERO, NOT/ONE, NOT/WIRE, NOT/NOT}\}$.

If NOT/NOT $\in P^{'}$, the NOT-Cell is obtained. Otherwise, if NOT/NOT $\notin P^{'}$, $P^{'} \subseteq \{\text{NOT/ZERO, NOT/ONE, NOT/WIRE}\}$. Assume $P^{'} = \{\text{NOT/ZERO, NOT/ONE, NOT/WIRE}\}$. Clearly, $P^{'}$ can not build NOT/NOT. Therefore, the NOT/NOT gate would always be built in the first step of the straightforward method.

Because *P* is complete, a circuit *C* which operates as NOT/NOT can be built by *P*. In the worst case, when all the combinations of the polymorphic gates in *P* are enumerated, the circuit *C* can be built. So, if *P* is complete, the straightforward method can build NOT/NOT. ∎

VI. THE THEORY AND METHOD FOR JUDGING THE COMPLETENESS OF POLYMORPHIC GATE SETS WITH MORE THAN TWO MODES

In this section, the theory and the straightforward method for judging the completeness of the polymorphic gate set with more than two modes are given.

*A. The Straightforward Method for Judging the Completeness of Polymorphic Gate Sets with More Than Two Modes*

In this subsection, the process to construct the AND-Cell is presented, and the process to construct the OR-Cell and NOT-Cell are similar to the AND-Cell.

For a polymorphic gate set $P = \{p_1, \cdots, p_n\} = \{g_{1,1}/g_{1,2}/\cdots/g_{1,m}, \cdots, g_{n,1}/g_{n,2}/\cdots/g_{n,m}\}$ and $p_i = g_{i,1}/g_{i,2}/\ldots/g_{i,m}$ ($1 \leq i \leq n$).

**Step 1.** Firstly, $\{g_{1,1}, g_{2,1}, \cdots, g_{n,1}, \text{logic-1}, \text{logic-0}\}$ is adopted to construct circuits $C_1^{(1)}, C_2^{(1)}, \cdots, C_{k^{(1)}}^{(1)}$. Each $C_i^{(1)}$ ($1 \leq i \leq k^{(1)}$) performs the AND function, and structures of $C_i^{(1)}$ and $C_j^{(1)}$ are different for any $1 \leq i < j \leq k^{(1)}$. Then, for each $g_{i,1}$ ($1 \leq i \leq n$) which is a building block of $C_j^{(1)}$, $g_{i,1}$ is replaced by $p_i$. So a polymorphic gate $PC_j^{(1)}$ is obtained which operates as a AND gate in mode 1. Therefore, a polymorphic gate set $P^{(1)} = \{PC_1^{(1)}, ..., PC_{k^{(1)}}^{(1)}\} = \{p_1^{(1)}, \cdots, p_{k^{(1)}}^{(1)}\} = \{\text{AND}/g_{1,2}^{(1)}/g_{1,3}^{(1)}/\cdots/g_{1,m}^{(1)}, \cdots, \text{AND}/g_{k^{(1)},2}^{(1)}/g_{k^{(1)},3}^{(1)}/\cdots/g_{k^{(1)},m}^{(1)}\}$ is obtained.

**Step 2.** $C_1^{(2)}, C_2^{(2)}, \cdots, C_{k^{(2)}}^{(2)}$ can be constructed by $\{g_{1,2}^{(1)}, g_{2,2}^{(1)}, \cdots, g_{k^{(1)},2}^{(1)}, \text{logic-1}\}$. Each $C_i^{(2)}$ ($1 \leq i \leq k^{(2)}$) performs the AND function, and structures of $C_i^{(2)}$ and $C_j^{(2)}$ are different for any $1 \leq i < j \leq k^{(2)}$. It should be noted that for any $1 \leq i \leq k^{(2)}$, if $g$ is a building block of $C_i^{(2)}$, the input of $g$ is never connected to logic-0. Similar to step 1, for any $1 \leq j \leq k^{(2)}$, the polymorphic circuit $PC_j^{(2)}$ is obtained from $C_j^{(2)}$ by replacing every $g_{i,2}^{(1)} \in C_j^{(2)}$ with $p_i^{(1)}$ for every $1 \leq i \leq k^{(1)}$. According to Lemma 1, $PC_j^{(2)}$ would operate the AND function in mode 1. So, the new obtained polymorphic gate set is $P^{(2)} = \{PC_1^{(2)}, \cdots, PC_{k^{(2)}}^{(2)}\} = \{p_1^{(2)}, \cdots, p_{k^{(2)}}^{(2)}\} = \{\text{AND}/\text{AND}/g_{1,3}^{(2)}/\cdots/g_{1,m}^{(2)}, \cdots, \text{AND}/\text{AND}/g_{k^{(2)},3}^{(2)}/\cdots/g_{k^{(2)},m}^{(2)}\}$.

.
.
.

**Step h.** $C_1^{(h)}, C_2^{(h)}, \cdots, C_{k^{(h)}}^{(h)}$ can be constructed by $\{g_{1,h}^{(h-1)}, g_{2,h}^{(h-1)}, \cdots, g_{k^{(h-1)},h}^{(h-1)}, \text{logic-1}\}$. Each $C_i^{(h)}$ ($1 \leq i \leq k^{(h)}$) performs the AND function, and structures of $C_i^{(h)}$ and $C_j^{(h)}$ are different for any $1 \leq i < j \leq k^{(h)}$. It should be noted that for any $1 \leq i \leq k^{(h)}$, if $g$ is a building block of $C_i^{(h)}$, the input of $g$ is never connected to logic-0. For any $1 \leq j \leq k^{(h)}$, the polymorphic circuit $PC_j^{(h)}$ is obtained by replacing every $g_{i,h}^{(h-1)} \in C_j^{(h)}$ with $p_i^{(h-1)}$ for every $1 \leq i \leq k^{(h-1)}$. According to Lemma 1, $PC_j^{(h)}$ would operate the AND function under mode from 1 to $h-1$. So, the obtained polymorphic circuit set $P^{(h)} = \{PC_1^{(h)}, ..., PC_{k^{(h)}}^{(h)}\} = \{p_1^{(h)}, \cdots, p_{k^{(h)}}^{(h)}\} = \{\text{AND}/\cdots/\text{AND}/g_{1,h+1}^{(h)}/\cdots/g_{1,m}^{(k)}, \cdots, \text{AND}/\cdots/\text{AND}/g_{k^{(h)},h+1}^{(h)}/\cdots/g_{k^{(h)},m}^{(h)}\}$.

.
.
.

The process should be done until the AND-Cell is generated, or this process gets stuck at step $h$ ($1 < h \leq m$), i.e. $\{g_{1,h}^{(h-1)}, g_{2,h}^{(h-1)}, \cdots, g_{k^{(h-1)},h}^{(h-1)}, \text{logic-1}\}$ can not construct a circuit operating as a AND gate.

The process to construct the OR-Cell and NOT-Cell is similar to the construction of the AND-Cell.

---

*Judge*($P, f, m$)

in: A polymorphic gate set $P = \{p_1, p_2, \cdots, p_n\} = \{g_{1,1}/\cdots/g_{1,m}, g_{2,1}/\cdots/g_{2,m}, \cdots, g_{n,1}/\cdots/g_{n,m}\}$ and $p_i = g_{i,1}/\cdots/g_{i,m}$ ($1 \leq i \leq n$). Parameter $f$ is AND or OR or NOT.

out: If $P$ can construct $f$, return true. Otherwise, return false.

1.   $d \leftarrow 1$
2.   $R \leftarrow \phi, I \leftarrow \phi$
3.   *stack* is initialized to empty
4.   **for** $i = 1$ to $m$ **do**
5.       $u_i \leftarrow 0$
6.   **do**{
7.       **if** $f \neq$ NOT
8.           $\{R, I\} \leftarrow$ *Construction_ANDOR*($P, f, d, R, I, m$)
9.       **else**
10.          $\{R, I\} \leftarrow$ *Construction_NOT*($P, d, R, I, m$)
11.       $u_d \leftarrow u_d + 1$
12.       **if** $R = \phi$ and $d = 1$ **then**
13.          **return** false
14.       **else**
15.          **if** $R = \phi$ **then**
16.             $d \leftarrow d - 1$
17.             $\{P, R, I\} \leftarrow$ *stack*.pop()
18.          **else**
19.             $d \leftarrow d + 1$
20.             *stack*.push{$P, R, I$}
21.             $P \leftarrow R, R \leftarrow \phi, I \leftarrow \phi$
22.   }**while**($d < m$ or $R = \phi$)
23.   **return** true

Fig. 14. *Judge*($P, f, m$) judges whether a polymorphic gate set $P$ can implement the function $f$. The routine "*Construction_ANDOR*(…)" and "*Construction_NOT*(…)" is given in Figure 6 and Figure 7, respectively. All polymorphic gates generated by $P$ at line 8 or line 10 form the set $I$. At line 8 and line 10, those polymorphic gates operating the function $f$ in the first $d$ modes form the set $R$. After the termination of the algorithm, the value of $u_d$ ($1 \leq d \leq m$) is the number of times the statements at line 8 or line 10 being operated, when the algorithm "*Construction_ANDOR*($P, f, d, \ldots$)" or "*Construction_NOT*($P, d, \ldots$)" works in mode $d$.

In Figure 14, the algorithm of the straightforward method for polymorphic gate sets with more than two modes is given. In each iteration of the Do-While loop (line 6 to line 22), a polymorphic gate set $R$ would be generated. And each polymorphic gate in $R$ would operate the function $f$ in the first $d$ modes. Therefore, when the algorithm terminates and $d = m$, a polymorphic gate which operates the function $f$ under all modes is obtained.

In the following part, an analysis of the time complexity of the algorithm in Figure 14 is given.

Firstly, the computing cost of the subroutine "*Construction_ANDOR*(…)" in Figure 6 is analyzed. It can be observed from line 7 to line 10 in Figure 6 that, the combinations of any three gates in the set *P_new* would be computed. Therefore, the computing cost of the loop started at line 7 is O($|P\_new|^3$). Suppose the Do-While loop is operated *loop* times, and in the $i^{th}$ loop, the polymorphic gate set "*P_new*" is denoted as $P^i$. Therefore, the computing cost of "*Construction_ANDOR*(…)" is $O(\sum_{i=1}^{loop}|P^i|^3)$. Similar to the analysis of "*Construction_ANDOR*(…)", the computing cost of "*Construction_NOT*(*P*, *d*, …)" in Figure 7 is $O(\sum_{i=1}^{loop}|P^i|^3)$ when *d* is 1, and $O(\sum_{i=1}^{loop}|P^i|^2)$ when *d* is greater than 1.

Secondly, the computing cost of "*Judge*(*P*, *f*, *m*)" in Figure 14 is analyzed. Without the loss of generality, suppose the input variable *f* of "*Judge*(*P*, *f*, *m*)" is AND or OR. Let's denote $C_{d,k}$ ($1\le d\le m$, $1\le k\le u_d$) the computing cost of "*Construction_ANDOR*(…)" at line 8, when "*Construction_ANDOR*(*P*, *f*, *d*, …)" is called the $k^{th}$ time in the working mode *d*. Therefore, the computing cost of "*Judge*(*P*, *f*, *m*)" is $O(\sum_{d=1}^{m}\sum_{k=1}^{u_d}C_{d,k})$. Clearly, $C_{d,k}$ is $O(\sum_{i=1}^{loop_{d,k}}|P_{d,k}^i|^3)$, where $loop_{d,k}$ ($1\le d\le m$, $1\le k\le u_d$) is the number of iterations of the Do-While loop in "*Construction_ANDOR*(*P*, *f*, *d*, …)" when it is called the $k^{th}$ time in the working mode *d*, and $P_{d,k}^i$ is the polymorphic gate set "*P_new*" in the $i^{th}$ Do-While loop of "*Construction_ANDOR*(*P*, *f*, *d*, …)" when it is called the $k^{th}$ time in the working mode *d*. Therefore, the computing cost of "*Judge*(*P*, *f*, *m*)" is $O(\sum_{d=1}^{m}\sum_{k=1}^{u_d}\sum_{i=1}^{loop_{d,k}}|P_{d,k}^i|^3)$.

In each iteration of the Do-While loop within "*Construction_ANDOR*(*P*, *f*, *d*, …)", at least one gate is inserted into the polymorphic gate set *P_new*. So, $|P_{d,k}^i|<|P_{d,k}^{i+1}|$, $|P_{d,k}^i|\le 16^{m-d}$ and $(\sum_{k=1}^{u_d}\sum_{i=1}^{loop_{d,k}}1)\le 16^{m-d+1}$.

Therefore, $\sum_{d=1}^{m}\sum_{k=1}^{u_d}\sum_{i=1}^{loop_{d,k}}|P_{d,k}^i|^3 < \sum_{d=1}^{m}(1^3+2^3+\cdots+16^{3\times(m-d+1)})$

$= \frac{1}{4}\times\sum_{d=1}^{m}(16^{4\times(m-d+1)}+2\cdot 16^{3\times(m-d+1)}+16^{2\times(m-d+1)})$

$= \frac{16^{4\times(m+1)}-16^4}{4\times(16^4-1)}+\frac{16^{3\times(m+1)}-16^3}{2\times(16^3-1)}+\frac{16^{2\times(m+1)}-16^2}{4\times(16^2-1)}$.

In the worst case, the time complexity of "*Judge*(*P*, AND, *m*)" and "*Judge*(*P*, OR, *m*)" is $O(16^{4\times m})$.

Similarly, when *f* is NOT, the computing cost is $O(\sum_{k=1}^{u_1}\sum_{i=1}^{loop_{1,k}}|P_{1,k}^i|^3+\sum_{d=2}^{m}\sum_{k=1}^{u_d}\sum_{i=1}^{loop_{d,k}}|P_{d,k}^i|^2)$. Therefore, in the worst case, the time complexity of "*Judge*(*P*, NOT, *m*)" is $O(16^{4\times m})$.

But, usually, the size of $P_{d,k}^i$ ($1\le d\le m$, $1\le k\le u_d$, $1\le i\le loop_{k,d}$) is not big, and the values of $u_d$ and $loop_{k,d}$ would be small. Therefore, the computing cost of the algorithm in Figure 14 is not high.

In [12], the algorithm for judging the completeness of a polymorphic gate set have to enumerate all the polymorphic signals that a polymorphic gate set can generate. In this paper, the straightforward method builds the AND-Cell, OR-Cell and NOT-Cell. And in fact, in most cases, with a relative low cost, the straightforward method can generate the three cells with a complete polymorphic gate set, especially for the manual version of the proposed method. For example, in the instance of Section VI.B, about ten polymorphic gates are built during the process of the straightforward method. And the total number of polymorphic gates with three modes is 4096.

*B. An Example*

The process to judge the completeness of {NAND/NOR/ANDNA, OR/ANDNB/XOR} is presented as an example.

**Example 5-1 {NAND/NOR/ANDNA, OR/ANDNB/XOR}**

Figure 15 illustrates the process to build the NOT-Cell. Firstly, the combination of {NAND/NOR, OR/ANDNB} can construct NOT/NOT, which is shown in Figure 15(*a*). Secondly, the NOT/NOT/ZERO gate can be gotten by replacing NAND/NOR and OR/ANDNB with NAND/NOR/ANDNA and OR/ANDNB/XOR, respectively. The structure of NOT/NOT/ZERO is shown in Figure 15(*b*). Thirdly, it is hoped that a different polymorphic gate which performs the NOT function in the first two modes can be obtained by modifying the structure of the circuit in Figure 15(*b*). The NOT-Cell is generated directly in Figure 15(*c*). The following steps give the process to construct the AND-Cell.

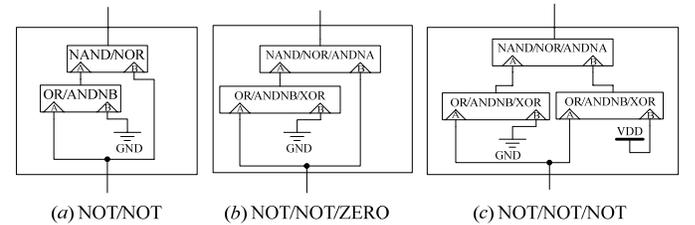

(*a*) NOT/NOT    (*b*) NOT/NOT/ZERO    (*c*) NOT/NOT/NOT

Fig. 15. The construction of the NOT-Cell.

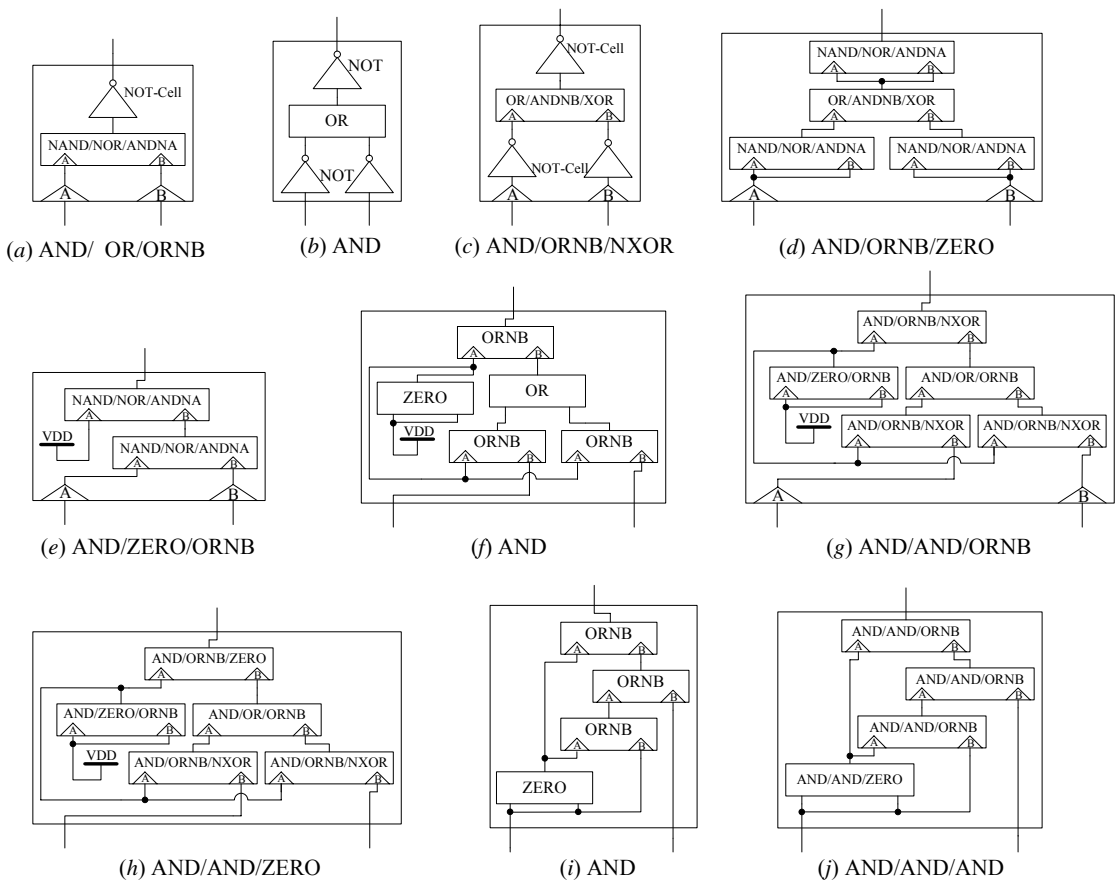

Fig. 16. The construction of the AND-Cell.

step 1. Because NOT(NAND($a$, $b$)) = AND($a$, $b$), AND/OR/ORNB depicted in Figure 16(*a*) can be generated by {NOT/NOT/NOT, NAND/NOR/ANDNA}. Figure 16(*b*) shows a way to build a AND gate from OR and NOT. So, AND/ORNB/NXOR and AND/ORNB/ZERO, shown in Figure 16(*c*) and Figure 16(*d*), can be obtained. Because NAND(1, NAND($a$, $b$)) = AND($a$, $b$), AND/ZERO/ORNB depicted in Figure 16(*e*) can be obtained. Therefore, a polymorphic gate set {AND/ORNB/NXOR, AND/ORNB/ZERO, AND/ZERO/ORNB, AND/OR/ORNB} is obtained.

step 2. A AND gate can be built by {ORNB, OR, ZERO}, which is shown in Figure 16(*f*). So, {AND/ORNB/NXOR, AND/ORNB/ZERO, AND/ZERO/ORNB, AND/OR/ORNB} can build AND/AND/ORNB and AND/AND/ZERO, which are shown in Figure 16(*g*) and Figure 16(*h*), respectively. Therefore, a polymorphic gate set {AND/AND/ORNB, AND/AND/ZERO} is obtained.

step 3. Because ORNB can build a AND gate depicted in Figure 16(*i*), {AND/AND/ORNB, AND/AND/ZERO} can build the AND-Cell shown in Figure 16(*j*).

According to De-Morgan rules, the OR-Cell can be obtained from the NOT-Cell and AND-Cell. Because the three cells can be constructed, according to Definition 1, {NAND/NOR/ANDNA, OR/ANDNB/XOR} is a complete polymorphic gate set.

### C. The Theory for Polymorphic Gate Sets with More Than Two Modes

Lemma 6, Lemma 7 and Lemma 8 prove the correctness of the straightforward method given in Section 5.2.

**LEMMA 6.** If $P = \{p_1, \cdots, p_n\} = \{g_{1,1}/g_{1,2}/\cdots/g_{1,m}, \cdots, g_{n,1}/g_{n,2}/\cdots/g_{n,m}\}$ ($m \geq 2$) is a complete polymorphic gate set, $P$ can construct the AND-Cell through the straightforward method.

**PROOF**. This lemma is proved by the induction on the number of modes $h$ of the polymorphic gate set $P$.

(1) $h = 2$. According to Lemma 3, $P$ can construct the AND-Cell through the straightforward method.

(2) For the induction step, assume the statement is true when $h \leq m$.

(3) $h = m + 1$.

Suppose $P'$ is a polymorphic gate set and $P' = \{p'_1, \cdots, p'_k\} = \{\text{AND}/g'_{1,2}/\cdots/g'_{1,m+1}, \cdots, \text{AND}/g'_{k,2}/\cdots/g'_{k,m+1}\} \cdot \{g'_{1,2}/\cdots/g'_{1,m+1}, \cdots, g'_{k,2}/\cdots/g'_{k,m+1}, \underbrace{\text{ZERO}/\cdots/\text{ZERO}}_{m}\}$ is complete. It is proved that $P$ can build such a polymorphic gate set $P'$.

Firstly, it is proved that such a polymorphic gate set $P'$ exists. For example, because $P$ is complete, $\{g_{1,2}/\cdots/g_{1,m}, \cdots, g_{n,2}/\cdots/g_{n,m}\}$ is a complete polymorphic gate set with $m - 1$ modes. Otherwise, if $\{g_{1,2}/\cdots/g_{1,m}, \cdots, g_{n,2}/\cdots/g_{n,m}\}$ is not complete, $\{g_{1,2}/\cdots/g_{1,m}, \cdots, g_{n,2}/\cdots/g_{n,m}\}$ can not build at least one of the three cells, i.e. AND-Cell, OR-Cell and NOT-Cell. Therefore, $P$ can not build at least one of the three cells. This contradicts that $P$ is complete. So, $\{g_{1,2}/\cdots/g_{1,m}, \cdots, g_{n,2}/\cdots/g_{n,m}\}$ is a complete polymorphic gate set. Let $P' = \{AND/g_{1,2}/\cdots/g_{1,m}, \cdots, AND/g_{n,2}/\cdots/g_{n,m}\}$, $P'$ is such a polymorphic gate set we wanted.

Secondly, it is proved that $P$ can build such a polymorphic gate set $P'$. Because $P$ is complete, for each $1 \leq i \leq k$, $P$ can construct a polymorphic circuit operating as $AND/g'_{i,2}/\cdots/g'_{i,m+1}$. That is to say, $P$ can build the polymorphic gate set $P'$.

Because $P$ is complete, $P$ can construct a polymorphic circuit operating as $AND/\underbrace{ZERO/\cdots/ZERO}_{m}$, and it is denoted as $C_{ZERO}$.

Because $\{g'_{1,2}/\cdots/g'_{1,m+1}, \cdots, g'_{k,2}/\cdots/g'_{k,m+1}, \underbrace{ZERO/\cdots/ZERO}_{m}\}$ is complete and the number of its modes is $m$. According to the induction assumption, $\{g'_{1,2}/\cdots/g'_{1,m+1}, \cdots, g'_{k,2}/\cdots/g'_{k,m+1}, \underbrace{ZERO/\cdots/ZERO}_{m}\}$ can construct a polymorphic circuit $C_{AND}$ operating as $\underbrace{AND/\cdots/AND}_{m}$. If each gate $g'_{i,2}/\cdots/g'_{i,m+1} \in C_{AND}$ is replaced with $p'_i$ for every $1 \leq i \leq k$, and $\underbrace{ZERO/\cdots/ZERO}_{m}$ is replaced with $C_{ZERO}$, a new polymorphic circuit $C'_{AND}$ is obtained from $C_{AND}$. Because logic-0 is not adopted in the construction of $C'_{AND}$, according to Lemma 1, $C'_{AND}$ would operate the AND function in mode 1. Therefore, $C'_{AND}$ would operate as $\underbrace{AND/\cdots/AND}_{m+1}$.

In summary, the statement is true when $h = m + 1$. ∎

**LEMMA 7.** If $P = \{p_1, \cdots, p_n\} = \{g_{1,1}/g_{1,2}/\cdots/g_{1,m}, \cdots, g_{n,1}/g_{n,2}/\cdots/g_{n,m}\}$ ($m \geq 2$) is a complete polymorphic gate set, $P$ can construct the OR-Cell through the straightforward method.

**LEMMA 8.** If $P = \{p_1, \cdots, p_n\} = \{g_{1,1}/g_{1,2}/\cdots/g_{1,m}, \cdots, g_{n,1}/g_{n,2}/\cdots/g_{n,m}\}$ ($m \geq 2$) is a complete polymorphic gate set, $P$ can construct the NOT-Cell through the straightforward method.

Lemma 7 can be proved similar to Lemma 6. Lemma 8 can be proved similar to Lemma 5.

## VII. DISCUSSION

In this paper, a straightforward method is proposed to judge the completeness of a polymorphic gate set. The proposed method is easy to understand and can be operated manually.

For a polymorphic gate set with $m$ modes, the judgment is done in $m$ steps. In the $k^{th}$ ($1 \leq k \leq m$) step, a polymorphic gate set is built by the polymorphic gate set obtained in the $(k - 1)^{th}$ step, and each new obtained polymorphic gate performs the AND (OR or NOT) function in the first $k$ modes. If the process finishes at the step $m$, the AND-Cell (OR-Cell or NOT-Cell) is generated. If all the three cells can be generated, according to Definition 1, the polymorphic gate set is complete. Both the manual version and the corresponding algorithm of the straightforward method are given. It is hard for the manual operation to enumerate all the combinations of gates. Therefore, the algorithm of the straightforward method is given. If a polymorphic gate set is complete, the algorithm returns true. Otherwise, it returns false.

Some heuristics can be added to accelerate the algorithm in Figure 14. For example, In Figure 6 and Figure 7, when the set $R$ is updated, check whether $R$ can build the function $f$ in each mode. If it is true, go to the next modes. Otherwise, continue the process in Figure 6 or Figure 7. This heuristic would reduce the number of backtrackings.

Compared with the method in [12], the method proposed in this paper has two advantages:

(1) It is straight forward and easy to understand. The judgment of a polymorphic gate set's completeness is done by constructing the AND-Cell, OR-Cell and NOT-Cell, and this process is similar to the completeness judgment of a traditional logic gate set[*] [14].

(2) It is suitable for manual operation. The proposed method consists of several independent steps. In each step, firstly, a traditional logic gate set is considered and it is adopted to build circuits which perform the AND or OR or NOT function. Secondly, those traditional logic gates are replaced with corresponding polymorphic gates, and polymorphic circuits are obtained. Some examples are given to show the process of the method. And these examples show that the straightforward method only generate a few polymorphic gates. For instance, in the example of Section VI.B, about ten polymorphic gates are built. Therefore, the actual computing cost of the proposed method is not high. And it is suitable for manually judging the completeness of polymorphic gate sets with two or three modes.

In [12], the strong and weak polymorphic gate sets are discussed. In this paper, the two kinds of polymorphic gate sets are not separately considered. However, it should be noted that if the straightforward method can construct the AND-Cell, OR-Cell and NOT-Cell without inputs of logic-0 and logic-1 by a polymorphic gate set, the gate set is strong complete. And if the straightforward method can construct the three cells with inputs of logic-0 and logic-1, the gate set is weak complete.

## VIII. CONCLUSION

In this paper, a straightforward method is proposed to judge the completeness of a polymorphic gate set. The steps to construct the AND-Cell, OR-Cell and NOT-Cell are given. If the three cells can be generated, the polymorphic gate set is complete. Some examples are presented to show the process of the construction of the three cells. The future work will study methods to design polymorphic circuits.

---

[*] If a traditional logic gate set can build {AND, OR, NOT}, it is a complete gate set.

ACKNOWLEDGMENT

This work is partly supported by the Fundamental Research Funds for the Central Universities.